# Discriminative Sleep Patterns of Alzheimer's Disease via Tensor Factorization

Yejin Kim[1], Xiaoqian Jiang[1], Luyao Chen[1], Xiaojin Li[2], Licong Cui[1]
[1] University of Texas Health Science Center, Houston, Texas, USA;
[2] Case Western Reserve University, Cleveland, Ohio, USA

**ABSTRACT**
Sleep change is commonly reported in Alzheimer's disease (AD) patients and their brain wave studies show decrease in dreaming and non-dreaming stages. Although sleep disturbance is generally considered as a consequence of AD, it might also be a risk factor of AD as new biological evidence shows. Leveraging National Sleep Research Resource (NSRR), we built a unique cohort of 83 cases and 331 controls with clinical variables and EEG signals. Supervised tensor factorization method was applied for this temporal dataset to extract discriminative sleep patterns. Among the 30 patterns extracted, we identified 5 significant patterns (4 patterns for AD likely and 1 pattern for normal ones) and their visual patterns provide interesting linkage to sleep with repeated wakefulness, insomnia, epileptic seizure, and etc. This study is preliminary but findings are interesting, which is a first step to provide quantifiable evidences to measure sleep as a risk factor of AD.

**INTRODUCTION**
Alzheimer's disease (AD) has become a major public health concern because of its increasing prevalence, chronicity, caregiver burden, and high personal and financial costs of care. About 25% to 40% of AD patients suffer from disturbances of sleep, such as insomnia at night, rapid eye movement (REM) sleep behavior disorder, agitated behavior at sunset and excessive sleeping during the daytime [1]. In addition, sleep apnea syndrome is prevalent in AD patients [2], which is associated with APOE4 that is a well-known risk factor of AD [3]. Those sleep problems occur early on in the course of AD, consistent with the finding that brain regions involved in sleep are affected during the development of AD. This fact implies that these sleep problems can be a preclinical marker for the development of AD [4]. A quantitative tool to measure such a sleep disturbance is electroencephalography (EEG), which is one of the most popular brain signals that records brain's spontaneous electrical activity on different regions of scalp over a period of time. This signal usually transforms to multiple waves according to frequency domain (e.g., Alpha, Theta, Delta). EEG can quantify sleep patterns. In the normal young adult, sleep consists of five cycle stages:

I. Stage 1 (5%): a period of transition from wakefulness (Alpha wave, 8-10.5 Hz) to light sleep (Theta wave, 4 - 8 Hz);
II. Stage 2 (50%): light sleep with Theta wave (4 - 8 Hz);
III. Stage 3 and Stage 4 (20%): slow-wave activity (SWA) or deep sleep with Delta wave (1 - 4 Hz);
IV. Stage 5 (25%): REM sleep with a fast desynchronized EEG containing Alpha (8 - 10.5 Hz), Beta (15 - 30 Hz), and Theta (4 - 8 Hz) waves [5].

With increasing age, sleep becomes "lighter" in that the percentage of Stage 1 increases and the percentage of Stage 3 and 4 decreases. Patients with AD show an increased number and duration of awakenings [6,7], consequently EEG recordings show that percentage of Stage 1 increases and percentage of Stage 2 and SWA (Stage 3 and 4) decreases more than normal aging does. Another interesting EEG pattern of AD is that the amount of REM sleep decreases and this change is mostly seen in later stages of AD [7]. Also, EEG slowing is observed prominently in REM sleep (other than in the awake EEG) [8,9]. These findings are driven by researchers who have been visually compared EEG between normal elderly and AD patients with their human naked eyes [9]. Unfortunately, this ad-hoc approach is not scalable to large samples and have difficulty in exploring a large combination of patterns associated with temporal changes on frequency bands and brain regions across sleep cycles [6]. Lack of quantifiable measurement can be another issue, which might lead to inconsistency in judgement. Therefore, a computational method that directly learns sleep patterns from observational data becomes more important in discovering meaningful and generalizable latent patterns of sleep EEG.

Dimensionality reduction is a widely-used method to discover such underlying latent space from high dimensional data. Particularly, linear models, such as principal component analysis, linear discriminant analysis, and nonnegative matrix factorization, represent the observed data as a weighted linear sum of latent dimensions and have been widely applied in EEG analysis [10]. Nonnegative tensor factorization (NTF) is another powerful linear model, which decomposes high dimensional data (such as time-frequency representation of EEG from multiple samples) into linear sum of basis components (which are interpreted as patterns). NTF has been applied to several medical domains, such as phenotyping [11,12] from electronic medical records, temporal patterns from behavioral log [13], and event-related EEG [14–16]. To derive discriminative phenotypes or patterns with respect to certain clinical outcome of interest, an extension of NTF, so called supervised NTF, has been proposed and enforces the basis components to be discriminative to the clinical outcome of interest [12].

In this work, we aim to discover temporal patterns from sleep EEG to differentiate AD from the control (non-AD), using supervised NTF. Previous studies on discovering AD patients' EEG patterns use only event-related EEG (not focused on sleep EEG, which capture very different characteristic of brain functionality). Also, they are unsupervised approach within AD populations (not supervised approach discriminating AD case and non-demented control) [17,18]. We focus on the sleep EEG and integrate the supervision term explicitly in our objective function to derive discriminative temporal patterns as *computational phenotypes*. Main contributions of our study are:

- *Novelty*: Our work is the first study to derive discriminative patterns of AD and classify AD using sleep EEG.
- *Secondary use of NSRR cohort:* We leveraged a rich collection of de-identified electrophysiological signals and clinical data elements (large cohorts) from National Sleep Research Resource (NSRR), which empowers retrospective secondary analyses.
- *Consistency:* Five representative patterns we derive were consistent with existing domain knowledge, which proves the validity of our method. We can further apply our validated methods to various applications.

**METHODS**
We developed a computational framework to discover sleep EEG patterns that would be discriminative to AD. This framework consists of three steps: i) patient matching; ii) EEG transformation; iii) tensor factorization (Figure 1).

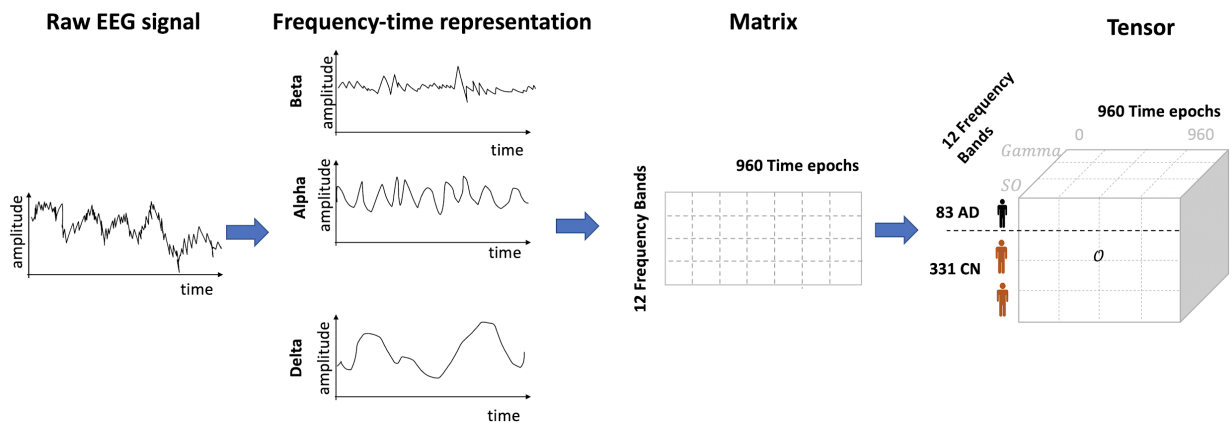

**Figure 1.** Overview of computational framework to discover discriminative patterns from Sleep EEG. This framework consists of three steps: i) patient matching; ii) EEG transformation; iii) tensor factorization. For each patient, the raw EEG signal was transformed into frequency-time representation using Fourier transform. The transformed EEG is a matrix format. We stacked all the individual matrices from each patient into a 3-order tensor.

**Datasets**
We leveraged two datasets in NSRR [19][20], which contains large collections of de-identified clinical data elements and electrophysiological signals from over 10 NIH-funded sleep cohort studies. Clinical data elements include demographic information (e.g., age, gender, race), anthropometric parameters (e.g., body mass index, height), physiologic measurements (e.g., diastolic/systolic blood pressure, heart rate), medical history (e.g., atrial fibrillation, cancer, depression), medications (e.g., antidepressant, acetylcholinesterase inhibitors for AD) and sleep symptoms (e.g., excessive daytime sleepiness). Electrophysiological signals in the European Data Format (EDF) are the polysomnography recordings (overnight sleep), with various channels such as electroencephalogram (EEG) and electrocardiogram (ECG). The two datasets we used were the Sleep Heart Health Study (SHHS) [21] and MrOS Sleep Study (MrOS) [22][23][24], where SHHS contains a total of 5,804 adults aged 40 or older, and MrOS contains a total of 2,911 men aged 65 years or older.

Since these two datasets are not specifically collected to study the relationship between sleep and AD, we performed retrospective analyses as a secondary use of these two datasets. To determine whether each patient has onset of AD we utilized a clinical variable on a specific medication, *acetylcholinesterase inhibitors*. *Acetylcholinesterase inhibitors* is widely used drug for AD. SHHS has a clinical variable asking if the patient has taken acetylcholinesterase inhibitors for AD within two weeks of the study visit. MrOS has three clinical data elements related to AD or AD medication use: (1) Has a doctor or other health care provider ever told you that you had dementia or AD? (2) Are you currently being treated for dementia or AD by a doctor? and (3) AD medication use.

**Cohort Selection using Propensity Score Matching**
Among 1,959 (MrOS) and 1,893 (SHHS) patients, we selected 331 controls normal (CN, non-AD) that match with 83 case (AD) with around 4:1 ratio in terms of potential confounding risk factors to AD. Due to complex nature of AD's neurodegeneration, many factors contribute to the disease. In order to focus on the relationship of AD and EEG signals, we need to reduce effects of other confounding variables that affect the incidence of AD, such as demographic background (e.g., age, gender, ethnicity) and potential risk factors (e.g., depression, hypertension, diabetes, cardiovascular disease, stroke). We utilized a statistical matching model called propensity score matching (PSM) [25], which matches estimators (i.e., confounders such as depression and hypertension) so that the distribution of EEG signals is independent of the outcome (i.e., AD), conditioned on the confounders [26]. The PSM algorithm finds the matched controls to the cases using two steps: i) computing propensity scores and ii) finding similar controls to each case using propensity scores. First, we obtained propensity scores as probability of AD computed from logistic regression, in which AD incidence is the binary label and confounders are predictors. Then we used *Radius matching* to find maximum *n* controls that fall within a predefined radius *r* (in terms of similarity of propensity scores). We conducted a line search to find the best combination of *n* and *r* that lead to biggest sample size within a tolerable bias.

**EEG Frequency Bands Representation**
We derived time-frequency representation from the raw EEG signals. EEG signals have various behaviors in different frequency bands, and the characteristics of different frequency bands are reported in previous work [27–29]. For example, previous studies demonstrated that spectral power is an important feature for sleep stage research [30], such as: the power of lower frequencies of EEG becomes stronger with the increasing depth of sleep; the EEG signal of Stage 1 has more power between 2-7 Hz; Stage 2 can be characterized by the presence of sleep spindles band (12-15 Hz), and SWA is defined when there are low frequency (less than 2 Hz) waves [31].

We first divided the entire sleep EEG records into non-overlap 30-second epochs during 8 hours (i.e., 8 hours / 30 seconds = 960 epochs). EGG signals after the 8 hours were truncated. After separating EEG time dimension into 960 epochs, we separated the EEG values into several waves depending on its frequency bandwidths using power spectral analysis. The most standard frequency bands include Delta wave, Theta wave, Alpha wave (Table 1). Fast Fourier transform (FFT) is a direct and commonly used spectral estimation method for the EEG frequency analysis [28]. The discrete Fourier transform (DFT) is an efficient numerical algorithm to perform Fourier transform in many practical applications. Let us denote *N* is the number of EEG values in one epoch and $x_n$ as a single EEG value in one epoch. The DFT of a sequence of *N* values $\{x_n\} := x_0, x_1, ..., x_{N-1}$ is defined as:

$$X_k = \frac{1}{N} \sum_{n=1}^{N} x_n \cdot e^{-(2\pi nk/N)i}$$

where $\{X_k\} := X_0, X_1, ..., X_{N-1}$ is a transformed sequence of $\{x_n\}$ for $k = 0, ..., n-1$, and *e* is the Euler's number. The power spectrum is then obtained with

$$P_k = \|X_k\|^2.$$

The $P_k$ values for $k = 0,\ldots,n-1$ are then separated according to the 12 frequency bands (Table 1). For example, $P_k$ values lying in every ¼ to ⅛ seconds (4 - 8 Hz) correspond to Theta wave. For each frequency band, we summed all the $P_k$ values and used it as an amplitude value. So, in total 12 frequency amplitude values are generated per every epoch.

**Table 1.** Extracted EEG frequency bands.

| Name | Hz | Name | Hz |
| --- | --- | --- | --- |
| Slow-Oscillations (SO) | [0.5, 1.5] | Sigma | [12.0, 15.0] |
| Slow-wave activity (SWA) | [0.5, 5.5] | Slow Sigma | [12.0, 13.5] |
| Delta | [1.0, 4.0] | Fast Sigma | [13.5, 15.0] |
| Theta | [4.0, 8.0] | Beta1 | [15.0, 20.0] |
| Alpha | [8.0, 10.5] | Beta2 | [20.0, 30.0] |
| Spindle | [10.5, 14.5] | Gamma | [30.0, 60.0] |

**Discovering Biomarkers using Tensor Factorization**

Using the EEG time-frequency representation of each patient, we constructed an observed tensor and extracted discriminative temporal patterns by applying supervised NTF to the observed tensor.

*Construct Tensor.* To derive patterns from time-frequency representation of EEG, we represent each patient's signal amplitude as a matrix (12 frequency bands × 960 time slots), see Figure 2. We applied log2-transform to the amplitude to make the distribution follow normal distributions and facilitate stable tensor factorization. We stacked those matrices into a third-order tensor $O$ with a shape of (83 AD + 331 CN) patients × 12 frequency bands × 960 time slots (Tensor is a generalization of matrix. Order of a tensor is the number of dimensions. A first-order tensor is a vector, a second-order tensor is a matrix, and tensors of order three or higher are called high-order tensors).

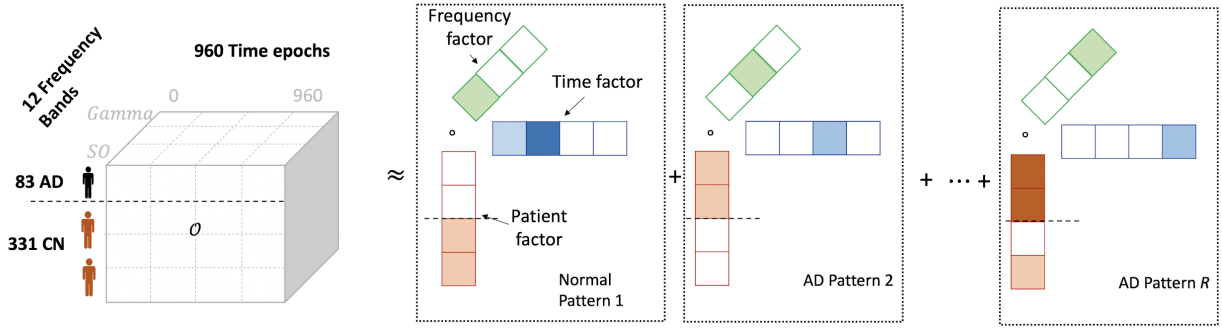

**Figure 2**. EEG time-frequency representation of all patients forms a tensor. AD= Alzheimer's disease, CN= Control.

*Regularized Nonnegative Tensor Factorization.* The most widely used tensor decomposition is CP method [32]. A third-order tensor $O$ with shape of $I \times J \times K$ is rank-one if it is an outer product of three vectors $a, b, c$, i.e., $O = a \circ b \circ c$ where $\circ$ means the vector outer product. $O_{ijk}$, the element at $(i, j, k)$ in the tensor $O$, is computed as product of elements in the vector, i.e., $O_{ijk} = a_i b_j c_k$. Tensor factorization (TF) is a dimensionality reduction approach that represents the original tensor as latent dimensions. The CP model approximates the original observed tensor $O$ as a linear combination of rank-one tensors [32]; that is, a third-order tensor $O$ is decomposed as minimizing difference between observed tensor and approximated tensor as

$$L = || O - \sum_{r=1}^{R} a_r \circ b_r \circ c_r ||^2$$

where a positive integer $R$ is the rank, $a_r, b_r, c_r$ are $r$-th column vectors in matrix $A, B, C$ with shape of $I \times R, J \times R, K \times R$, respectively. Here, $A, B, C$ are called as factor matrices. When tensor $O$ contains non-negative data (such as amplitude, intensity or counts), we set non-negative constraints $A, B, C \geq 0$ for interpretability of latent dimensions, which is so called nonnegative tensor factorization (NTF). Since our objective is to derive discriminative patterns, we used supervised version of TF, which adds a supervised regularizer that encourages the patterns to be separated according to AD vs CN [12]. The supervised TF adds logistic regression regularizer as

$$L - \mu \cdot \log P(A, y|\theta) = L - \mu \cdot \frac{1}{1 + exp(-y \cdot \theta A)}$$

where $\mu$ is a weight parameter to balance the tensor error and the loss on regularizer, $\theta$ is parameter for logistic regression, and $y$ is label ($y = 1$ if AD; -1 CN). To further enhance interpretability via compact patterns, we also added $l_1$ norm regularizer to the factor matrix. The $l_1$ regularizer shrinks the less important coefficients to zero, as improving interpretability of model:

$$L + \lambda \cdot (||A||_1 + ||B||_1 + ||C||_1)$$

where $\lambda$ is a weight parameter to balance the tensor error and the $l_1$ norm loss. Therefore, our final objective function is adding both of regularizers:

$$L - \mu \cdot \log P(A, y|\theta) + \lambda \cdot (||A||_1 + ||B||_1 + ||C||_1).$$

*Discover Patterns using NTF.* We applied this regularized NTF to our EEG dataset. We let factor matrix *A, B*, and *C* represent patients (AD + CN), frequency bands, and time slots, respectively. The *R* latent dimensions in factor matrices *A, B,* and *C* represent temporal patterns of EEG. Each EEG pattern consists of a set of frequency bands and a set of time slots. Individual frequency band or time slot contributes to the pattern with different extent of membership, and the amount of contribution is stored in columns of *B* and *C*. Likewise, individual patients have characteristics of the *R* patterns with different extent of membership. That is, the column vector of *B* defines how much the frequency bands are involved in each pattern, and the column vector of *C* defines how much the time slots are involved in each pattern. The row vector of *A* defines how much patients participate (or have membership) in the characteristic of each pattern.

**EXPERIMENTS AND RESULTS**

**Propensity Score Matching**
To create our training cohort for regularized NTF, we obtained controls for the cases from our data sources (MrOS, SHHS). We used propensity score matching to reduce the effect of age, race, common risk factors (hypertension, depression, stroke, diabetes) while trying to keep as much cases as possible. After matching, we lost a few cases in order to reduce the bias. We ended up with 83 cases and 331 controls at a roughly 1:4 ratio. Tables 2a and 2b summarize the statistics on features before and after the matching.

**Table 2a:** Statistics of MrOS dataset before and after matching

| Feature | Original Dataset | | PS Matched Dataset | |
|---|---|---|---|---|
| | AD cohort | Control cohort | AD cohort | Control cohort |
| **Number of Patients** | 86 | 2,397 | 55 | 203 |
| **Age** | 80.14±5.36 | 77.6±5.53 | 78.87±5.57 | 78.21±5.59 |
| **Race** | 86% | 88.6% | 94.5% | 98.0% |
| **Hypertension** | 50% | 81.6% | 67.3% | 78.8% |
| **Depression** | 5.8% | 3.7% | 3.6% | 2.0% |
| **Stoke** | 10.5% | 6.3% | 3.6% | 1.0% |
| **Diabetes** | 14.0% | 18.7% | 14.5% | 6.9% |

**Table 2b:** Statistics of SHHS dataset before and after matching

|  | Original Dataset | | PS Matched Dataset | |
|---|---|---|---|---|
| **Feature** | **AD cohort** | **Control cohort** | **AD cohort** | **Control cohort** |
| **Number of Patients** | 33 | 1,860 | 28 | 126 |
| **Age** | 79.6±6.9 | 73.5±7.83 | 78.9±7.1 | 76.5±6.67 |
| **Gender - Female** | 48.5% | 53.5% | 50% | 42.9% |
| **Race - Caucasian** | 87.9% | 88.9% | 89.3% | 98.4% |
| **Race - African American** | 9.1% | 8.8% | 7.1% | 1.6% |
| **Cardiovascular Disease** | 39.4% | 29.2% | 39.3% | 23.8% |
| **Hypertension** | 51.5% | 95.6% | 60.7% | 96.8% |
| **Stoke** | 6.1% | 4.5% | 0% | 1.6% |

**Discovering Biomarker using Tensor Factorization**

We implemented the regularized NTF using Pytorch 11.4 with adaptive momentum estimation (ADAM) for optimization. We set the maximum number of iterations as 1,000. The running time was less than 15 secs with 3 parallel GPUs. We added dropout to logistic regression coefficients for robustness. After extensive parameter tuning, we set dropout rate=0.5, $R$=30, $\mu = 0.1$, and $\lambda = 0.05$.

*Evaluating NTF methods.* We computed discriminative power, sparsity, overlap of patterns that are derived from the NTF methods. We measured the discrimination by the area under the receiver operating characteristic curve (AUC) to classify AD and control. We measured compactness by sparsity and overlap of the temporal patterns. High sparsity means a few frequency bands or time period dominantly characterize each pattern whereas the other bands or time period are negligible, making interpretation of the patterns easy. The sparsity was computed as an averaged Gini index of involvement values in each pattern (i.e., the column vectors of *B* and *C*) [33]. The overlap measures the degree of overlapping between all pattern pairs [12]. Patterns with less overlap are more distinctly identified. The overlap is computed as an averaged cosine similarity between all pair of column vectors of *B* and *C*. We also computed mean squared error (MSE) to evaluate how closely the derived patterns reflect the observed original data. We computed mean and standard deviation after ten repeated trials. We compared the discriminative power and compactness with different settings of regularizers:

- NTF: Basic NTF model without any regulariziers ($\mu = 0, \lambda = 0$)
- NTF + logit: NTF with supervised regularizer based on logistic regression likelihood ($\mu = 0.1, \lambda = 0$)
- NTF +$l_1$norm: NTF with $l_1$norm ($\mu = 0, \lambda = 0.05$)
- NTF +$l_1$norm+logit: NTF with $l_1$norm and logistic regression likelihood ($\mu = 0.1, \lambda = 0.05$)

Table 3 summarizes interested measurements. We found that NTF with $l_1$norm and supervised term outperforms other baselines in terms of discriminative power and compactness. The NTF +$l_1$norm+logit showed the highest AUC, sparsity, and lowest overlap (Table 3). The $l_1$norm regularizer improved compactness (i.e., increased sparsity and decreased overlap). The supervised regularizer also improved discriminative power (i.e., increased AUC).

**Table 3**. Discrimination and compactness comparison. We computed average and standard deviation after 10 repeated trials. The number of phenotypes $R$=30.

| Methods | AUC | Sparsity | Overlap | MSE |
| --- | --- | --- | --- | --- |
| NTF ($\mu = 0, \lambda = 0$) | 0.6716 (0.0321) | 0.1799 (0.0078) | 0.8936 (0.0094) | 0.3061 (0.0038) |
| NTF + $l_1$norm ($\mu = 0, \lambda = 0.05$) | 0.6684 (0.0385) | 0.1821 (0.0105) | 0.8914 (0.0111) | 0.3068 (0.0023) |
| NTF + logit ($\mu = 0.1, \lambda = 0$) | 0.6743 (0.0328) | 0.1788 (0.0069) | 0.8956 (0.0081) | 0.3076 (0.0034) |
| NTF + $l_1$norm+logit ($\mu = 0.1, \lambda = 0.05$) | 0.6852 (0.0259) | 0.1822 (0.0066) | 0.8913 (0.0075) | 0.3079 (0.0025) |

*Evaluating Individual Patterns.* We presented patterns that an NTF +$l_1$norm+logit model derives (with AUC=0.6981). After deriving $R$=30 patterns, we learned a logistic regression model using the $R$=30 patterns as predictors and selected patterns that separate AD and CN with statistical significance (Table 4: Coefficient). Five patterns showing $p$-value < 0.05 were: Pattern 2 for CN and Patterns 4, 18, 28, and 30 for AD. That is, Patterns 4, 18, 28, and 30 were positively related to AD whereas Pattern 2 was negatively related to AD. To verify this relationship, we reported the number of patients from AD and CN according to the extent of involving to each pattern, the membership values (Table 4: Membership distribution). We found that in AD patterns (4, 18, 28, and 30) the ratio of AD patients to CN increases as the membership value increases. In contrast, in CN pattern 2 the ratio of AD patients to CN decreases as the membership value increases. These findings are consistent with the logistic regression results in which AD patients are more likely to have larger values on the Patterns 4, 18, 28, and 30; and to have smaller values on the Pattern 2. Note that we initially made cohort with ratio of AD:CN=1:4, thus the number of CN was always larger than the number of AD across all membership values.

We visualized the five representative patterns (Table 4: Visualization). We represented each pattern using heatmap, according to the membership values of frequency bands or time slots. Frequency bands or time slots that dominantly characterize the pattern showed high values (bright yellow), whereas frequency bands or time slots that are less involved in the pattern showed low values (dark purple). Slow oscillation, slow wave activity, and spindle were overlapped with Delta and Sigma waves, so we denoted them separately where each band belongs to.

Patterns can be interpreted in details as follows:
- *Pattern 2 - Healthy normal sleep.* Pattern 2 was more prevalent to non-AD group. Pattern 2 forms a slow oscillation, slow wave activity and spindle, which none of AD EEG patterns have. Pattern 2 is consistent with the existing known fact that AD patient rarely forms spindle in Sigma wave (12 - 15 Hz) [34].
- *Pattern 4 and 18 - Epileptic seizure during sleep.* Pattern 4 refers to Delta and Theta wave (0.5 Hz - 8 Hz) with slow oscillation and slow wave activity and also Gamma wave (30 Hz ~). This activity occurs during at the beginning and the end of sleep. Similarly, Pattern 18 refers to Theta wave (4 - 8 Hz) with Gamma wave (30 Hz ~) during the middle of sleep (4 - 7 hour). Because Gamma wave are not usually seen during sleep, we hypothesized that this combination of slow wave and high frequency wave is due to epilepsy, which is very common in AD and also main cause of AD [35]. A Gamma wave increase is found in epileptic patients, probably reflecting both cortical excitation and perceptual distortions such as deja vu phenomenon frequently observed in epilepsy [36]. During an epileptic seizure the EEG are mainly sharp waves and spikes that may also appear as spike-wave complexes in combination with the slow waves [36].
- *Pattern 28 - drowsy sleep.* Pattern 28 was focused on Theta wave 4-8 Hz and Alpha 8-10.5 Hz waves several times throughout the 8 hours of sleep. This pattern is consistent with typical AD patient's sleep pattern. It is known that AD patients usually wake up several times during nighttime [37]. As a result, the percentage of wakefulness and Stage 1 increases, which refer to the transition of the brain from Alpha waves (8-10.5 Hz) to Theta waves (4-8 Hz).
- *Pattern 30 - Insomnia.* Pattern 30 has very strong signals on Beta and Gamma waves (15 Hz ~) during the first one hour and 5 - 8 hour of sleep. This sleep EEG pattern is accordant with that of insomnia patients, in which Beta and Gamma activity usually increase [38]. Insomnia is one of common comorbidities of AD patients [1].

**Table 4**. Summary of representative five patterns. Logistic regression coefficient and *p*-value to classify AD from CN. Membership values distribution of AD and CN, and visualization of each pattern. SO= slow oscillation, SWA=slow wave activity. Range of membership values varies according to pattern. SO, SWA, and spindle were presented separately due to overlap.

| EEG pattern | Coefficient (*p*-value) | Membership distribution | Visualization (Dominant frequency bands and time period) |
|---|---|---|---|
| 2 (CN) | -2.7164 (0.034) | 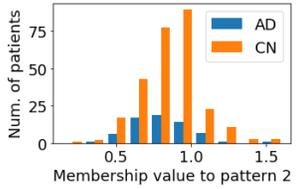 | 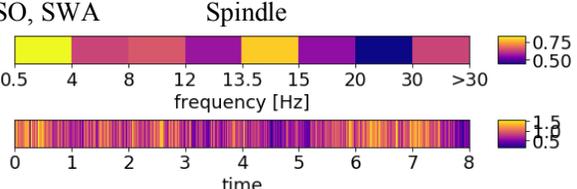 |
| 4 (AD) | 1.8200 (0.013) | 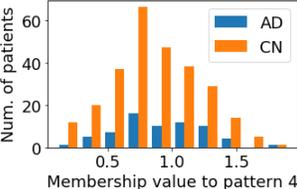 | 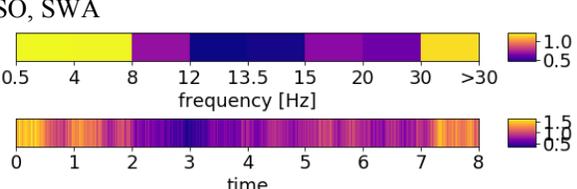 |
| 18 (AD) | 1.4873 (0.034) | 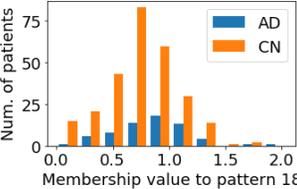 | 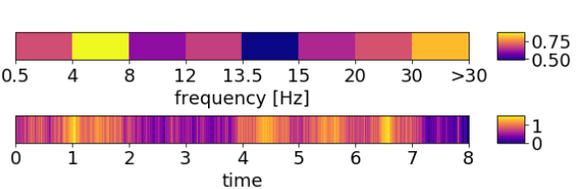 |
| 28 (AD) | 2.4115 (0.028) | 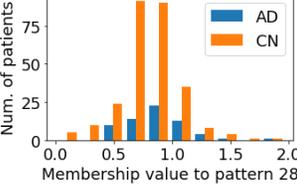 | 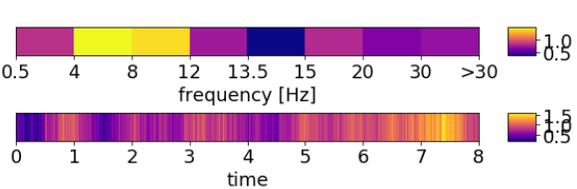 |
| 30 (AD) | 2.1408 (0.013) | 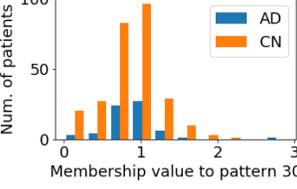 | 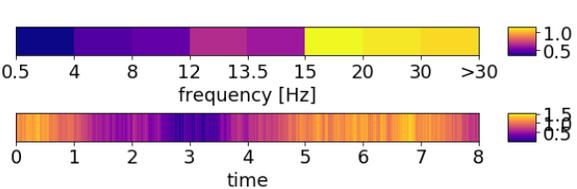 |

**LIMITATION AND CONCLUSION**

Data collected in National Sleep Research Resource (NSRR) was meant for study sleeping. Although many clinical variables are collected, they are mostly focused on sleep but not specific to the AD population. Based on the populational AD occurrence rate (at different ages), we believe there are more eligible patients for the study but it is hard to confirm. So, we end up with a relatively small population but results are still very interesting as we identified several distinctive patterns that are significantly associated with AD. These results still need further verification and biological explanation. Another limitation is related to the onset time of AD.

In this study, we aimed to derive sleep patterns of AD patients using EEG signals. From National Sleep Research Resource, we built a unique cohort of 83 cases and 331 controls with clinical variables and EEG signals. We used a regularized nonnegative tensor factorization that can derive discriminative patterns. Among the 30 patterns extracted, we identified five significant patterns with $p <0.05$: Health normal sleep with spindle; epileptic seizure during sleep; drowsy sleep with repeated wakefulness; and insomnia. This study is preliminary but finding are interesting, which is a first step to provide quantifiable evidences to measure sleep as a risk factor of AD.

**REFERENCES**
1   Ju Y-ES, Lucey BP, Holtzman DM. Sleep and Alzheimer disease pathology—a bidirectional relationship. Nature Reviews Neurology. 2014;**10**:115–9. doi:10.1038/nrneurol.2013.269
2   Bliwise DL, Tinklenberg J, Yesavage JA, *et al.* REM latency in Alzheimer's disease. Biological Psychiatry. 1989;**25**:320–8. doi:10.1016/0006-3223(89)90179-0
3   Kadotani H. Association Between Apolipoprotein E ∈4 and Sleep-Disordered Breathing in Adults. JAMA. 2001;**285**:2888. doi:10.1001/jama.285.22.2888
4   Postuma RB, Gagnon JF, Vendette M, *et al.* Quantifying the risk of neurodegenerative disease in idiopathic REM sleep behavior disorder. *Neurology* 2009;**72**:1296–300.
5   Petit D, Gagnon J-F, Fantini ML, *et al.* Sleep and quantitative EEG in neurodegenerative disorders. Journal of Psychosomatic Research. 2004;**56**:487–96. doi:10.1016/j.jpsychores.2004.02.001
6   Prinz PN, Vitaliano PP, Vitiello MV, *et al.* Sleep, EEG and mental function changes in senile dementia of the Alzheimer's type. *Neurobiol Aging* 1982;**3**:361–70.
7   Dykierek P, Stadtmüller G, Schramma P, *et al.* The value of REM sleep parameters in differentiating Alzheimer's disease from old-age depression and normal aging. Journal of Psychiatric Research. 1998;**32**:1–9. doi:10.1016/s0022-3956(97)00049-6
8   Hassainia F, Petit D, Montplaisir J. Significance probability mapping: The final touch int-statistic mapping. Brain Topography. 1994;**7**:3–8. doi:10.1007/bf01184832
9   Tsolaki A, Kazis D, Kompatsiaris I, *et al.* Electroencephalogram and Alzheimer's Disease: Clinical and Research Approaches. International Journal of Alzheimer's Disease. 2014;**2014**:1–10. doi:10.1155/2014/349249
10  Parra LC, Spence CD, Gerson AD, *et al.* Recipes for the linear analysis of EEG. NeuroImage. 2005;**28**:326–41. doi:10.1016/j.neuroimage.2005.05.032
11  Kim Y, Sun J, Yu H, *et al.* Federated Tensor Factorization for Computational Phenotyping. *KDD* 2017;**2017**:887–95.
12  Kim Y, El-Kareh R, Sun J, *et al.* Discriminative and Distinct Phenotyping by Constrained Tensor Factorization. *Sci Rep* 2017;**7**:1114.
13  Choi J, Rho MJ, Kim Y, *et al.* Smartphone dependence classification using tensor factorization. *PLoS One* 2017;**12**:e0177629.
14  Mørup M, Hansen LK, Herrmann CS, *et al.* Parallel Factor Analysis as an exploratory tool for wavelet transformed event-related EEG. NeuroImage. 2006;**29**:938–47. doi:10.1016/j.neuroimage.2005.08.005
15  Lee H, Kim Y-D, Cichocki A, *et al.* Nonnegative tensor factorization for continuous EEG classification. *Int J Neural Syst* 2007;**17**:305–17.
16  Hunyadi B, Dupont P, Van Paesschen W, *et al.* Tensor decompositions and data fusion in epileptic electroencephalography and functional magnetic resonance imaging data. Wiley Interdisciplinary Reviews: Data Mining and Knowledge Discovery. 2017;**7**:e1197. doi:10.1002/widm.1197
17  Latchoumane C-FV, Vialatte F-B, Solé-Casals J, *et al.* Multiway array decomposition analysis of EEGs in Alzheimer's disease. *J Neurosci Methods* 2012;**207**:41–50.
18  Spyrou L, Parra M, Escudero J. Complex Tensor Factorization With PARAFAC2 for the Estimation of Brain Connectivity From the EEG. *IEEE Trans Neural Syst Rehabil Eng* 2019;**27**:1–12.
19  Dean DA, Goldberger AL, Mueller R, *et al.* Scaling Up Scientific Discovery in Sleep Medicine: The National Sleep Research Resource. Sleep. 2016;**39**:1151–64. doi:10.5665/sleep.5774

20  Zhang G-Q, Cui L, Mueller R, *et al.* The National Sleep Research Resource: towards a sleep data commons. Journal of the American Medical Informatics Association. 2018;**25**:1351–8. doi:10.1093/jamia/ocy064
21  Group SHHR, Sleep Heart Health Research Group, Redline S, *et al.* Methods for Obtaining and Analyzing Unattended Polysomnography Data for a Multicenter Study. Sleep. 1998;**21**:759–67. doi:10.1093/sleep/21.7.759
22  Blank JB, Cawthon PM, Carrion-Petersen ML, *et al.* Overview of recruitment for the osteoporotic fractures in men study (MrOS). *Contemp Clin Trials* 2005;**26**:557–68.
23  Orwoll E, Blank JB, Barrett-Connor E, *et al.* Design and baseline characteristics of the osteoporotic fractures in men (MrOS) study--a large observational study of the determinants of fracture in older men. *Contemp Clin Trials* 2005;**26**:569–85.
24  Blackwell T, Yaffe K, Ancoli-Israel S, *et al.* Associations Between Sleep Architecture and Sleep-Disordered Breathing and Cognition in Older Community-Dwelling Men: The Osteoporotic Fractures in Men Sleep Study. Journal of the American Geriatrics Society. 2011;**59**:2217–25. doi:10.1111/j.1532-5415.2011.03731.x
25  Rosenbaum PR. The Central Role of the Propensity Score in Observational Studies for Causal Effects. In: *Matched Sampling for Causal Effects*. Cambridge University Press 2006. 170–84.
26  Rampichini LGC. Propensity scores for the estimation of average treatment effects in observational studies. https://www.bristol.ac.uk/media-library/sites/cmm/migrated/documents/prop-scores.pdf (accessed 2 Jan 2019).
27  Greene RW, Frank MG. Slow Wave Activity During Sleep: Functional and Therapeutic Implications. The Neuroscientist. 2010;**16**:618–33. doi:10.1177/1073858410377064
28  Thakor NV, Tong S. Advances in Quantitative Electroencephalogram Analysis Methods. Annual Review of Biomedical Engineering. 2004;**6**:453–95. doi:10.1146/annurev.bioeng.5.040202.121601
29  Tamminen J, Payne JD, Stickgold R, *et al.* Sleep spindle activity is associated with the integration of new memories and existing knowledge. *J Neurosci* 2010;**30**:14356–60.
30  Fraiwan L, Lweesy K, Khasawneh N, *et al.* Automated sleep stage identification system based on time–frequency analysis of a single EEG channel and random forest classifier. Computer Methods and Programs in Biomedicine. 2012;**108**:10–9. doi:10.1016/j.cmpb.2011.11.005
31  Li X, Cui L, Tao S, *et al.* HyCLASSS: A Hybrid Classifier for Automatic Sleep Stage Scoring. IEEE Journal of Biomedical and Health Informatics. 2018;**22**:375–85. doi:10.1109/jbhi.2017.2668993
32  Douglas Carroll J, Chang J-J. Analysis of individual differences in multidimensional scaling via an n-way generalization of 'Eckart-Young' decomposition. *Psychometrika* 1970;**35**:283–319.
33  Hurley N, Rickard S. Comparing measures of sparsity. In: *2008 IEEE Workshop on Machine Learning for Signal Processing*. 2008. doi:10.1109/mlsp.2008.4685455
34  Montplaisir J, Petit D, Lorrain D, *et al.* Sleep in Alzheimer's Disease: Further Considerations on the Role of Brainstem and Forebrain Cholinergic Populations in Sleep-Wake Mechanisms. Sleep. 1995;**18**:145–8. doi:10.1093/sleep/18.3.145
35  Born HA. Seizures in Alzheimer's disease. Neuroscience. 2015;**286**:251–63. doi:10.1016/j.neuroscience.2014.11.051
36  Herrmann C, Demiralp T. Human EEG gamma oscillations in neuropsychiatric disorders. Clinical Neurophysiology. 2005;**116**:2719–33. doi:10.1016/j.clinph.2005.07.007
37  Silber MH, Ancoli-Israel S, Bonnet MH, *et al.* The visual scoring of sleep in adults. *J Clin Sleep Med* 2007;**3**:121–31.
38  Perlis ML, Smith MT, Andrews PJ, *et al.* Beta/Gamma EEG activity in patients with primary and secondary insomnia and good sleeper controls. *Sleep* 2001;**24**:110–7.